\documentclass[aps,prl,reprint,twocolumn,showpacs,floatfix,superscriptaddress]{
revtex4-1}

\usepackage{amssymb,amsmath,amstext}                
\usepackage{graphicx}                                               
\usepackage{epstopdf}                                               
\usepackage{color}                                                     
\usepackage{bm}                                                        
\usepackage{appendix}                                              
\usepackage[utf8]{inputenc}
\usepackage{latexsym}
\usepackage{xcolor}
\usepackage{braket}
\usepackage{ulem}
\usepackage{siunitx}
\usepackage{umoline}
\usepackage{epsfig}
\usepackage{graphics}
\usepackage{amssymb}
\usepackage{psfrag}
\usepackage{mathtools}
\usepackage[english]{babel}
\usepackage[T1]{fontenc}
\usepackage{lmodern}
\usepackage{booktabs}

\usepackage{rotating}
\usepackage{multirow}

\definecolor{lblue} {RGB}{51,71,158}
\usepackage[colorlinks=true,citecolor=blue,linkcolor=blue,urlcolor=lblue]{hyperref}

\begin{document}

\title{Polynomially 
filtered exact diagonalization approach to many-body localization}

\author{Piotr Sierant}
\affiliation{Institute of Theoretical Physics, Jagiellonian University in Krakow, \L{}ojasiewicza 11, 30-348 Krak\'ow, Poland }
\email{piotr.sierant@uj.edu.pl}
\affiliation{ICFO- Institut de Sciences Fotoniques, The Barcelona Institute of Science and Technology, Av. Carl Friedrich Gauss 3, 08860 Castelldefels (Barcelona), Spain}
\author{Maciej Lewenstein}
\affiliation{ICFO- Institut de Sciences Fotoniques, The Barcelona Institute of Science and Technology, Av. Carl Friedrich Gauss 3, 08860 Castelldefels (Barcelona), Spain}
\affiliation{ICREA, Pg. Lluís Companys 23, 08010 Barcelona, Spain}
\email{maciej.lewenstein@icfo.eu}
\author{Jakub Zakrzewski}
\affiliation{Institute of Theoretical Physics, Jagiellonian University in Krakow, \L{}ojasiewicza 11, 30-348 Krak\'ow, Poland }
\affiliation{Mark Kac Complex
Systems Research Center, Jagiellonian University in Krakow, \L{}ojasiewicza 11, 30-348 Krak\'ow,
Poland. }
\email{jakub.zakrzewski@uj.edu.pl}

\date{\today}

\begin{abstract}
Polynomially filtered exact diagonalization method (POLFED) for large sparse matrices is introduced. 
The algorithm finds
an optimal basis of a subspace spanned by eigenvectors with eigenvalues close to a specified energy target by a spectral transformation 
using a high order polynomial of the matrix.
The memory requirements scale better with system size than in the
state-of-the-art shift-invert approach. 
The potential of POLFED is demonstrated  examining many-body localization transition in
1D interacting quantum spin-1/2 chains. We investigate the
disorder strength and system size scaling of Thouless time. System size 
dependence of bipartite entanglement entropy and of the gap 
ratio highlights the importance of finite-size effects. We  discuss possible scenarios regarding 
the many-body localization transition obtaining estimates for the critical disorder strength. 
\end{abstract}

\maketitle

{\it Introduction.} Qantum many-body systems are generically expected to approach equilibrium
according to eigenstate thermalization hypothesis \cite{Deutsch91, Srednicki94, Alessio16}. The 
phenomenon of many-body localization (MBL) \cite{Nandkishore15, Alet18, Abanin19} provides a robust class   of
many-body systems which fail to reach thermal equilibrium  \cite{Oganesyan07, Pal10, Kjall14,
Lev16, Mondaini15, Prelovsek16, Sierant17, Kozarzewski18, Sierant18, Mace19}.
Further examples of non-ergodic behavior include Stark localization \cite{Schulz19,vanNieuwenburg19}, 
persistent oscillations \cite{Turner18,Wen19, Khemani19, Iadecola19, Iadecola19a},
the presence of confinement \cite{James19,Chanda20b}, Hilbert 
space fragmentation \cite{Sala20,Khemani19b, Rakovszky20} or lack of 
thermalization in lattice gauge theories \cite{Smith17, Brenes18, Magnifico19, 
Chanda20, Giudici19, Surace19}.

Classification of many-body systems according to their ergodic properties 
is a fascinating new direction of research, however, it poses  serious 
technical challenges as exact methods are restricted either to small system 
sizes \cite{Pietracaprina18} or allow to trace time evolution only within a 
short time interval \cite{Enss17, Doggen18, Chanda19}. Hence, a 
fully consistent theory of MBL transition is missing, with recent 
approaches pointing towards Kosterlitz-Thouless scaling 
\cite{Goremykina19, Morningstar19, Dumitrescu19,Laflorencie20, Suntajs20}.
The finite-size effects strongly influence exact diagonalization (ED) results, 
{leading to a recent debate \cite{Suntajs19, Sierant20b, Abanin19a, Panda19} about}
discriminating between
finite size effects and asymptotic features 
of disordered many-body systems.

The example of MBL transition shows that development of ED techniques allowing 
to study thermalization properties of possibly large many-body systems is 
in demand. 
In this letter, we introduce a polynomially filtered exact 
diagonalization (POLFED) as a tool to calculate eigenvectors of large sparse matrices with eigenvalues 
close to a specified energy target.  The polynomial spectral 
transformation preserves the sparse structure of matrices 
avoiding the main bottleneck of shift-invert method of exact diagonalization (SIMED) \cite{Pietracaprina18}.
We employ POLFED in study of MBL transition in disordered quantum spin chains 
unveiling new aspects of system size scaling of Thouless time, entanglement 
entropy and level statistics.  Our results provide novel qualitative and quantitative arguments
in favor of the existence of MBL transition in the thermodynamic limit.

{\it Benchmark models. }
We consider 1D disordered spin chains with Hamiltonian: 
\begin{equation}
 \hat{H}=\sum_{l=1}^2  \sum_{i=1}^{L} J_l\left( 
S^x_{i}S^x_{i+l}+S^y_{i}S^y_{i+l} + \Delta S^z_{i}S^z_{i+l}  \right) + 
\sum_{i=1}^{L} h_i S^z_i,
 \label{eq: XXZ}
\end{equation}
where  $\vec{S}_i$ are spin-1/2 matrices, $L$ is the system size, $J_1=1$ is 
fixed as the energy unit,
periodic boundary conditions are assumed and $h_i \in [-W, W]$ 
are independent, uniformly distributed random variables.
The XXZ model, widely studied in the MBL context \cite{Agarwal15, Bera15, 
Bera17, Herviou19, 
Colmenarez19, Sierant20}, is obtained for $J_2 = 0$ and $\Delta = 1$. The choice $J_2=1$ 
and $\Delta=0.55$
leads to the $J_1$-$J_2$ model studied in \cite{Suntajs19}.
The Hamiltonian \eqref{eq: XXZ} becomes a real symmetric sparse matrix 
$H\in \mathbb{R}^{\mathcal N \times \mathcal N}$ in basis 
of eigenstates of $S^i_z$ operator; the matrix size, $\mathcal 
N$, in the zero magnetization $\sum_i S^z_i =0$ sector is given by 
$\mathcal N = {L\choose{L/2}} \propto e^{L \ln 2}/\sqrt{L}$.

  \begin{figure}
 \includegraphics[width=0.99\linewidth]{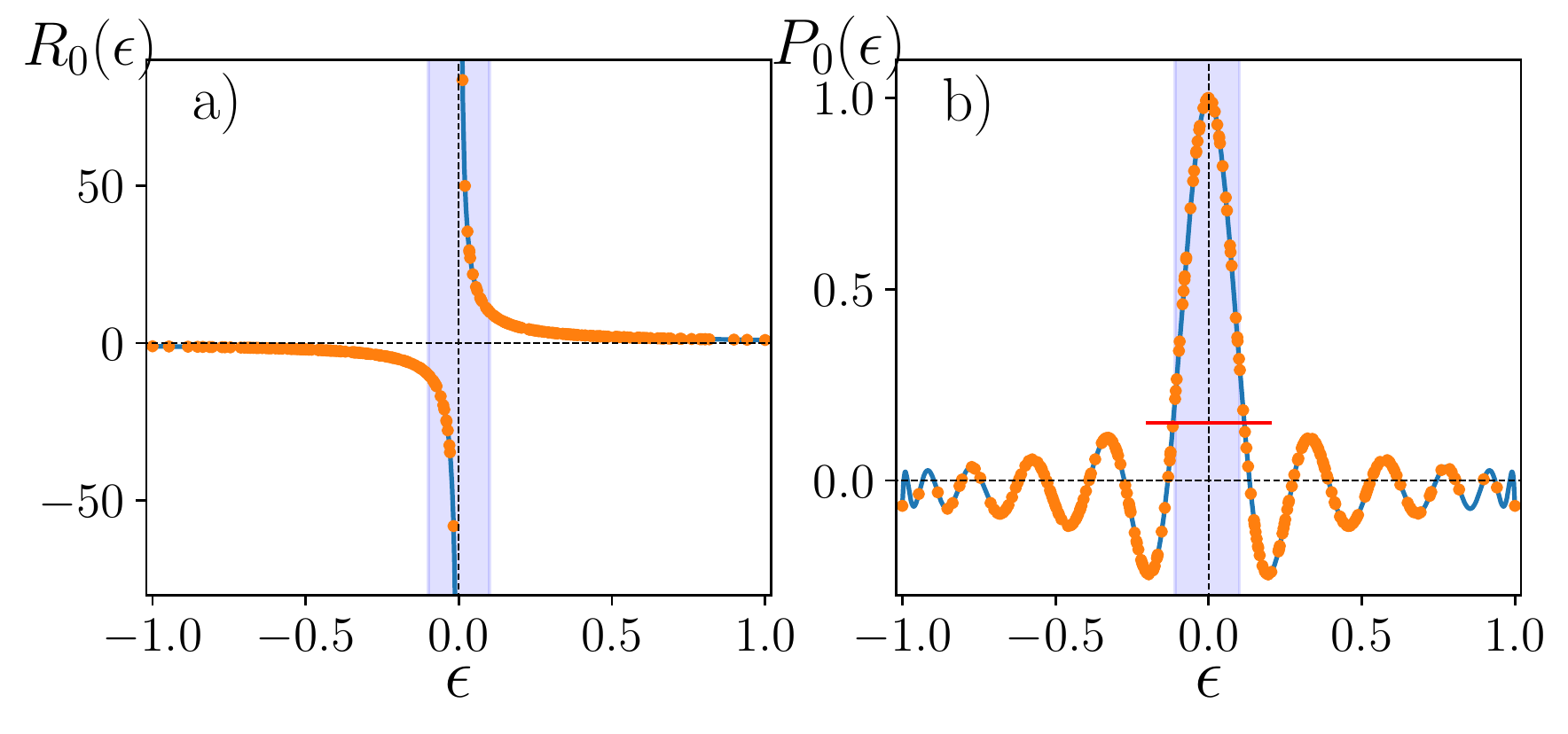} 
  \caption{ Spectral transformation employed in a) SIMED; b) 
POLFED algorithm. 
The spectrum is transformed 
according to a) $R_0(\epsilon)$; b) $P^{K=22}_{\,\sigma=0}(\epsilon)$.
Eigenvectors corresponding to eigenvalues at the edges of the transformed 
spectrum (shaded areas) are accessible for iterative methods.
 }\label{methodIll}
\end{figure}  

{\it Calculation of eigenpairs.} 
Hamiltonians of many-body systems
are typically characterized by exponential scaling of matrix size, $\mathcal{N}$, with the system size, $L$, and
sparsity in appropriately chosen basis.
For a sparse matrix the number of non-zero 
entries, $N_{nz}$, is much smaller than $\mathcal N^2$ implying that matrix 
vector multiplication requires much less operations than for a dense 
matrix. The Lanczos algorithm \cite{Lanczos50} utilizes this fact to 
find exterior  eigenpairs {(corresponding to
highest/lowest eigenvalues)}.
However, due to an increasing density of states and reorthogonalization 
costs, Lanczos algorithm becomes inefficient if many eigenpairs are requested. 
In contrast, a full ED procedure \cite{Golub12} allows one to determine all eigenpairs of $H$ 
but, with present day computers, it is limited to $\mathcal N \lesssim 5\cdot 10^4 $ corresponding to 
$L=18$ in \eqref{eq: XXZ}.
Larger matrix sizes are tractable by SIMED \cite{Pietracaprina18}.
The Hamiltonian is transformed via $H \rightarrow R_{\sigma}(H) = 
(\sigma-H)^{-1}$ so that eigenvalues close to $\sigma$ become exterior 
eigenvalues of the matrix $ R_{\sigma}(H)$, see Fig.~\ref{methodIll}.
Consequently, the Laczos algorithm for 
the matrix $R_{\sigma}(H)$ converges to eigenpairs close to 
the target $\sigma$.
The Lanczos iteration 
with $R_{\sigma}(H)$ is performed by calculating $LU$ decomposition 
\cite{Amestoy01, Amestoy06} of the matrix $H$.
That has a significant drawback:
the sparsity pattern of $H$ is lost resulting in a very severe for large 
$\mathcal N$ phenomenon of fill-in of the matrix. This
was identified as the main bottleneck of SIMED when
applied to quantum many-body systems \cite{Pietracaprina18}.

{\it POLFED algorithm.}
To avoid the fill-in phenomenon and utilize the sparsity of the $H$ matrix in 
an efficient way, we use the \textit{polynomial spectral 
transformation}
\begin{eqnarray}
H \rightarrow P^K_{\sigma}(H) = \frac{1}{D}\sum_{n=0}^K c^{\sigma}_n T_n(H)
\label{polyTransf}
\end{eqnarray}
where $T_n(x)$ denotes $n$-th Chebyshev polynomial, 
the coefficients $c^{\sigma}_n = \sqrt{4-3 \delta_{0,n}} \cos(n \arccos{\sigma})$ are 
obtained from expanding a Dirac delta function centered at $\sigma$ in 
Chebyshev polynomials and normalization $D$ assures that $P_{\sigma}( \sigma 
)=1$. The eigenvalues close to the target energy $\sigma$ are the largest 
eigenvalues of the transformed matrix $P_{\sigma}(H)$ as shown in 
Fig.~\ref{methodIll}b). Hence, a block Lanczos method \cite{Cullum74, Golub77} 
applied to matrix $P^K_{\sigma}(H)$ converges to 
eigenpairs close to the target $\sigma$.
We note that eigensolvers employing polynomial spectral 
transformations were considered also in \cite{Bekas08, Fang12, Li16diag, Pieper16}.

The POLFED consists of the 
following steps. Lanczos algorithm is used 
to find the lowest (highest) eigenvalue $E_0$ ($E_1)$ of matrix $H$ which is 
then rescaled to $\tilde{H} = [2H-(E_0+E_1)]/(E_1-E_0)$.
The order $K$ of 
transformation \eqref{polyTransf} is specified by requiring that the number 
of eigenvalues $\theta_i$ of $P^K_{\sigma}(\tilde H)$ accessible to 
Lanczos algorithm (belonging to the shaded area in Fig.~\ref{methodIll})
is equal to a number of requested eigenvalues $N_{ev}$ -- as the condition 
we take 
{$\theta_i \geqslant p = 0.17$}.
To find the value of $K$, an 
estimate of density of states $\tilde{\rho}(\sigma)$ at energy $\sigma$ of the 
matrix $\tilde{H}$ is needed. The $\tilde{\rho}(\sigma)$ can be found 
efficiently for arbitrary sparse matrices using iterative methods 
\cite{Silver94, Silver96}. For the benchmark models \eqref{eq: XXZ}, 
the density of states is Gaussian and is well approximated by an analytic 
expression $\tilde{\rho}(0)= (E_1-E_0)\mathcal{N}/\Gamma$ at the center of 
spectrum $\sigma=0$ where $\Gamma \propto \sqrt{L}W$. Having found $K$, the 
POLFED algorithm, starting with a matrix of orthonormalized random vectors $Q_1
\in \mathbb R^{\mathcal N \times s}$,  performs the block Lanczos iteration 
\begin{eqnarray}
\label{Lan1}
U_j = P^K_{\sigma}(\tilde H) Q_j - Q_{j-1} B^T_{j},  \quad A_j = Q_j^T U_j  \\
\label{Lan2}
R_{j+1} = U_j - Q_j A_j, \quad Q_{j+1} B_{j+1} = R_{j+1}, 
\end{eqnarray}
where $Q_0=0$, $B_0=0$ and the second operation in \eqref{Lan2} is $QR$ decomposition. The 
iteration is repeated for $j=1,\ldots,m$ resulting in $Q_j, U_j, R_j \in 
\mathbb R^{\mathcal N \times s}$ and $A_j, B_j \in  \mathbb R^{s \times s}$ 
matrices. 
In exact arithmetic, columns of $Q_j$ matrices form an orthonormal set of 
vectors. This property is gradually lost with increasing $m$ during 
calculations with a finite precision. Hence, between \eqref{Lan1} and \eqref{Lan2}, we perform a re-orthogonalization  of columns 
of matrix $U_j$ against the columns of matrices $ \{Q_i\}_{i=1}^j$.
The product of $P^K_{\sigma}( \tilde H)$ with each column of $Q_j$ in 
\eqref{Lan1} is computed with the Clenshaw algorithm \cite{Clenshaw55}.
The orthogonal matrix $\mathcal Q_m=[Q_1,\ldots,Q_m] \in \mathbb 
R^{\mathcal N \times ms} $ {defines a block tridiagonal matrix 
$T_m = \mathcal Q_m^T P^K_{\sigma}( \tilde H) \mathcal Q_m$}
with $A_j$ matrices on the diagonal 
and $B_j$ ($B_j^T$) below (above) the diagonal.
The eigenvectors $t_i \in \mathbb{R}^{ms}$ of $T_m$ are used to calculate 
$u_i = \mathcal Q_m t_i$ which converge, with increasing $m$, to
exterior eigenvectors of $P^K_{\sigma}( \tilde H)$ \cite{Saad80}, that is to
eigenvectors of $\tilde H$ with eigenvalues close to the target
$\sigma$. The convergence is reached after $m$ 
steps when the residual norm $|| B_{j+1} \tilde{t}_i ||$ \cite{Golub77}
(where  $\tilde{t}_i$ are the last $s$ components of the vector $t_i$ and $|| u || = 
\sqrt{u^T u} $) vanishes within the numerical precision for each eigenvector $t_i$ corresponding 
to eigenvalue $\theta_i \geqslant p$. 
 The eigenvalues of the 
matrix $\tilde H$ are found as $\varepsilon_i = u_i^T \tilde H u_i$ and the 
convergence is verified by a direct calculation of the residual norms $|| 
\tilde H u_i - \varepsilon_i u_i ||$. 
Each of our tests shows that eigenvalues $\varepsilon_i$ are, within 
numerical precision, equal to $N_{ev}$ eigenvalues of $\tilde H$ 
closest to the target $\sigma$.
For technical details of the algorithm see \cite{suppl}.

The POLFED is tailored for maximal efficiency in calculations for 
quantum many-body systems. The order $K$ of the polynomial transformation 
\eqref{polyTransf} scales linearly with the density of states $ \tilde{\rho}(\sigma)$
that increases exponentially with system size $L$. Thus, the product $P^K_{\sigma}(\tilde H) Q_j$ in \eqref{Lan1}
is the most time consuming step of the calculation.
POLFED offers high scalability as the product can be parallelized in two manners: i) it splits into 
independent multiplications of subsequent columns of $Q_j$ 
by $P^K_{\sigma}(\tilde H)$; ii) each of the matrix vector multiplications can be 
parallelized. The re-orthogonalization step between \eqref{Lan1} and 
\eqref{Lan2} can be parallelized in a similar manner. 
The number $m$ of iterations after which the algorithm converges is proportional 
to $N_{ev}$. Hence, the memory consumption, dominated by 
$\mathcal Q_m$, scales as $N_{ev} \mathcal N $.
 The memory requirements of SIMED are larger and scale as $c(L)\mathcal N$
where the factor $c(L)$ is due to the fill-in of the matrix. 
For $XXZ$ model $c(L)\propto 3^{L/2}$ \cite{Pietracaprina18}. Moreover,
$c(L)$ grows rapidly with number $N_{nz}$ of non-zero elements of the matrix
significantly increasing the resources needed in calculations for $J_1$-$J_2$ model.
In contrast, computation time of POLFED increases linearly with $N_{nz}$ -- resources
for $XXZ$ and $J_1$-$J_2$ models are comparable, for detailed benchmarks see \cite{suppl}. 
POLFED allows to find larger number of eigenpairs in a single run than the recently proposed
eigensolver \cite{Beeumen20}. This reduces fluctuations of averages over eigenstates
and is essential in calculation of the Thouless time.

{\it Thouless time. } The spectral form factor is defined as $K(\tau)  =  \langle | 
\sum_{j=1}^{\mathcal N} g( E_j) \mathrm{e}^{-i E_j \tau} |^2  \rangle/Z$, where $E_j$ are eigenvalues of 
$H$ after an unfolding procedure \cite{Gomez02}, 
$g(\epsilon)$ is a Gaussian function, the average is taken over
disorder realizations and $Z$ is a normalization constant assuring $ K(\tau) \stackrel{\tau\rightarrow\infty}{\rightarrow}1$.
The spectral form factor of many-body system (with time reversal invariance) 
follows Gaussian Orthogonal Ensemble (GOE) prediction $K(\tau)=K_{GOE}(\tau)$ only for $\tau > 
\tau_{Th}$ defining the Thouless time $t_{Th} = \tau_{Th} t_H$, where 
$t_H=2\pi \rho(0)$ is the Heisenberg time.

 \begin{figure}
 \includegraphics[width=0.99\linewidth]{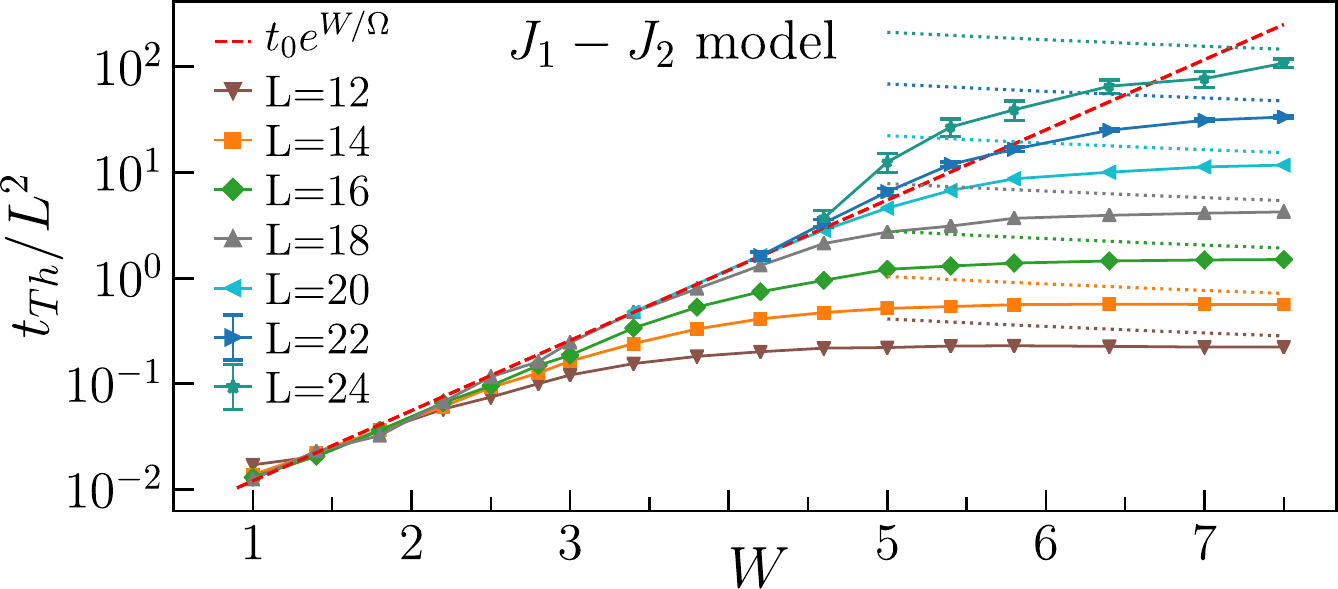} 
  \caption{ Thouless time $t_{Th}$ for system size $L$ and disorder strength $W$ for 
$J_1$-$J_2$ model. The dotted lines denote Heisenberg time $t_H$; the dashed line denotes
  a scaling $t_{Th} \propto L^2 \mathrm{e}^{W/\Omega}$ broken by the
  $L=22, 24$ data.
 }\label{Thoul}
\end{figure}  

The Thouless time, $t_{Th}$, calculated for $J_1$-$J_2$ spin chains of length 
$L\leqslant18$ \cite{Suntajs19} scales as
$t_{Th} \propto L^2 \mathrm{e}^{W/\Omega}$ where $W$ is the disorder strength and 
$\Omega$ is constant. 
If this scaling prevailed in $L\rightarrow \infty$ limit, it would imply $t_{Th}/t_H 
\rightarrow 0$ so that the 
system would be well described by GOE and MBL phase would be absent 
for arbitrary disorder strength in the thermodynamic limit.
To verify this surprising conclusion we supplement results of full ED 
of $J_1$-$J_2$ model 
with Thouless times obtained with POLFED for $L=20,22,24$
respectively for $800$, $200$, $50$ disorder realizations.
Since we calculate $N_{ev}=2500$ eigenvalues in the middle of spectrum ($\sigma=0$),
the sum in the definition of spectral form factor $K(\tau)$ is truncated.
However, this does not influence the value of $t_{Th}$ as long as it is larger than a certain threshold value 
determined by $N_{ev}$ \cite{suppl}. The obtained Thouless times are shown in Fig.~\ref{Thoul}.
Data for $L\leqslant20$ follows the scaling $t_{Th} \propto L^2 \mathrm{e}^{W/\Omega}$ deviating from it at disorder 
strength $\tilde W(L)$ which increases with the system size, for instance $\tilde W(18)\approx 3.7$ or $\tilde 
W(20)\approx4.6$. This behavior changes qualitatively for $L=22, 24$ data breaking the scaling
$t_{Th} \propto L^2 \mathrm{e}^{W/\Omega}$. Similar behavior heralds Anderson localization
transition in single particle disordered systems \cite{Sierant20b}, hence, our data suggest the presence of  the
transition to MBL phase in $J_1$-$J_2$ model. 
Therefore, one has to reach a sufficiently large $L$ to see the correct scaling of Thouless time, which
raises the question about the finite size effects at MBL transition.

  \begin{figure*}
 \includegraphics[width=0.99\linewidth]{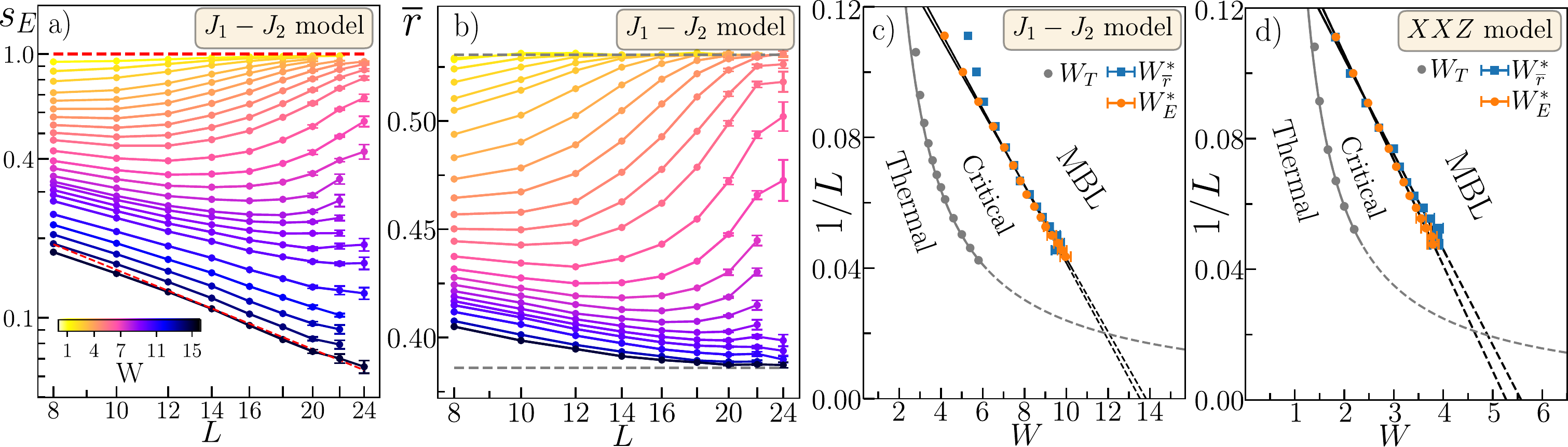} \vspace{-0.3cm}
  \caption{ Finite size effects at MBL transition. a) The entanglement entropy $s_E$
  of eigenstates of $J_1$-$J_2$ model vs system size $L$ 
for disorder strengths  $W=1.4, ..., 15$ (denoted on the color bar), dashed lines correspond to ergodic and 
MBL behavior; b) the same for the gap ratio $\overline r$, dashed lines correspond to GOE and 
Poisson limits;
 c) $W^*_{E, \overline r}$ and $W_T$  as function $1/L$  for $J_1$-$J_2$ model (see text); d) the same for 
$XXZ$ model.
 }\label{figCOMB}
\end{figure*}  

{\it Entanglement entropy and level statistics. } 
The entanglement entropy 
allows for insights in nature of MBL transition
\cite{Yu16, Khemani17, Khemani17a}. The entanglement 
entropy of an eigenstate
is defined
as $S_E=-\sum_i \alpha_{i}^2 \log(\alpha_{i}^2)$, where $\alpha_{i}$ are Schmidt basis coefficients (see e.g. 
\cite{Karol}) associated with the bipartition of the lattice into subsystems containing sites $[x, x+L/2)$ and  
$[x+L/2, x+L)$ (the sites are numbered modulo $L$). We average $S_E$ over the position of the cut $x$, over 
$N_{ev} \leqslant \min\{\mathcal N/100, 2000\}$ eigenstates in the middle of the spectrum
($\sigma=0$) of $J_1$-$J_2$ model for system sizes $12\leqslant L \leqslant 24$ (for $L=8,10$ we take 
$N_{ev}=5$) as well as over 
more than $5000$, $200$, $50$ disorder realizations respectively for $L\leqslant20$, $L=22$, $L=24$.
Finally, we obtain the scaled entanglement entropy $s_E=S_E/S_{RMT}(L)$ where $S_{RMT}(L)=(L/2) \ln(2) + 
(1/2 + \ln(1/2))/2-1/2$ corresponds to a chaotic spin chain in the total $\sum_i S^z_i=0$ 
sector \cite{Vidmar17}. The resulting $s_E$ is shown in Fig.~\ref{figCOMB}a). For available 
system sizes, the scaled entanglement entropy $s_E$: i) monotonically 
increases with $L$ for $W \lesssim 3.4$; ii) monotonically decreases for $W\gtrsim 11 $; iii) 
decreases for {smaller $L$}
and starts increasing for larger system sizes (a similar reentrant behavior 
was observed e.g. in \cite{Serbyn15, Panda19}).
The behavior i) clearly leads to an ergodic system at large $L$. In contrast, for large disorder strengths 
e.g. $W=15$, an area law of entanglement entropy \cite{Bauer13,Serbyn13a}  $s_{E} \propto 1/L$  {arises due to} 
the emergent integrability of MBL phase \cite{Serbyn13b,Huse14, Ros15,Imbrie16, Wahl17, Mierzejewski18, Thomson18}. 
Averaging $r_i=\min \{g_{i},g_{i+1} \} / \max\{g_{i},g_{i+1}\} $ (where 
$g_i=E_{i+1}-E_{i}$) over eigenvalues corresponding to eigenstates from which $s_E$ was calculated,
we obtain a mean gap ratio $\overline r$ shown in Fig.~\ref{figCOMB}b).
The mean gap ratio $\overline r$ probes level statistics of the system, admitting values characteristic
for GOE and Poisson statistics for ergodic and localized systems  \cite{Oganesyan07, Atas13}. 
Similarly as for $s_E$, the mean gap ratio $\overline r$ follows the three types of behavior with system size depending 
on disorder strength $W$.

To understand whether and at which disorder
strength the MBL transition takes place one has to study the interplay between the ii) and iii) {trends}.
To this end, 
we find the disorder strength $W^*_E{(L)}$  such that 
$s_E(L-1)=s_E(L+1)$ for odd $L$ and $s_E(L-2)=s_E(L+2)$ for even $L$, for details see 
\cite{suppl}. 
Smooth changes of $s_E$ with $L$ and $W$ assure that
$W^*_E(L)$ is the largest disorder strength, for a given system size $L$, at which
the volume-law  $S_E\propto L$, expected
for an ergodic system, is still obeyed. 
Consequently, the 
disorder strength $W^*_E(L)$ is a lower bound for the critical disorder strength $W_C$ of the transition to MBL 
phase.
Fig.~\ref{figCOMB}c) shows  the relation between $W^*_E$ and $1/L$
along with disorder strength $W^*_{\overline r}$ 
obtained in analogous manner for the average gap ratio $\overline r$. 
Another aspect of finite size effects at MBL transition is revealed when, for given $L$, one finds
a disorder strength $W_T(L)$ for which the scaled entanglement entropy is close to the ergodic limit, e.g.
$s_E(W_T)=0.8$. Such a criterion yields $W_T\propto L$. 
Equivalently, $W_T$ can be found as a disorder strength for which the average gap 
ratio $\overline r$ departs from the GOE limit \cite{Suntajs19}.
This allows us to identify the following regimes:
A) thermal, for $W < W_T$, with entanglement entropy fulfilling the volume-law and
close to the value for chaotic spin chain $S_E \approx S_{RMT}(L)$
and level statistics well described by GOE; B) critical, for $W^*_{E, \overline r} < W < W_T $, with 
$S_E < S_{RMT}(L)$ but scaling super linearly with $L$ and value of $\overline r$ increasing with $L$ towards 
the GOE limit; C) MBL, for $W < W^*_{E, \overline r}$, with both scaled entanglement entropy $s_E(L)$ 
and average gap ratio $\overline r$  decreasing with system size $L$.
Fig.~\ref{figCOMB}d) shows that behavior of the $XXZ$ model is similar
(data for $s_E$ and $\overline r$ can be found in \cite{suppl}). 
The three regimes resemble the 
qualitative picture of MBL transition proposed in \cite{Khemani17a}.

The asymptotic features of disordered spin chains depend on how  $W^*_{E, \overline r}$ and $W_T$
behave in thermodynamic limit. For available system sizes, $8 \leqslant L \leqslant 24$,  
the linear scaling of $W_T$ with $L$ as well as the linear scaling of $W^*_{E, \overline r}$
with inverse of system size $1/L$, denoted by solid lines in Fig.~\ref{figCOMB} c), d), 
are accurately obeyed. Extrapolating the scalings (dashed lines in the same Fig.),
leads to the crossing $W_T=W^*_{E, \overline r}$ at $L_0\approx 50$ showing the incompatibility 
of the two scalings. Thus, it seems conceivable that studying eigenstates at 
at system size $L_0$ would yield conclusive results about the $L\rightarrow \infty$ limit,
c.f. \cite{Panda19}. However, it is also possible that either of the 
scalings breaks down at smaller $L$ achievable in the near future with POLFED.

The unveiled linear dependence of $W^*_{E, \overline r}$
on $1/L$ is consistently approached by data for all system sizes.
Extrapolating to $L\rightarrow \infty$ limit, we get estimates of critical disorder strength
\begin{eqnarray}
W^{J_1-J_2}_C \approx 13.7 \quad \quad \mathrm{and} \quad \quad W^{XXZ}_C \approx 5.4,
  \label{WCs}
\end{eqnarray}
respectively for $J_1$-$J_2$ and $XXZ$ models.
Our estimate for $W^{XXZ}_C$ is larger than the value
$W_C\approx 3.7$ for XXZ model \cite{Luitz15} (which  yields the critical exponent violating the 
 Harris criterion \cite{Harris74, Chayes86, Chandran15a}) or the estimates obtained
after an asymmetric scalings on both sides of the transition:
$W_C\approx 3.8$ \cite{Mace18}, $W_C\approx 4.2$ \cite{Laflorencie20}.
Since our approach relies on an analysis of the drift of crossing points of $s_E(W)$ and
$\overline r(W)$ curves,
it does not rely on any finite size scaling procedure. 
Our estimate for
$W^{XXZ}_C$ is consistent with the lower bound $W_C>4.5$ of \cite{Devakul15} 
as well as with
$W_C > 5$ obtained in analysis of quench 
dynamics of large XXZ spin chain \cite{Doggen18}.

{\it Conclusions. } 
The POLFED algorithm, thanks to the employed polynomial spectral transformation,
has a better scaling of computation time with matrix size $\mathcal N$ than 
the state-of-the-art SIMED algorithm.
Avoiding the fill-in phenomenon, POLFED has a significantly lower memory consumption than SIMED, 
moreover, its performance decreases only linearly with increasing the number of non-zero off-diagonal
matrix entries. For those reasons POLFED opens new pathways in studies of highly excited states of many-body systems with
potential applications to
systems with long-range interactions realized in experiments with
polar molecules \cite{Yan13}, Rydberg atoms \cite{Browaeys20}, trapped ions \cite{Richerme14, Jurcevic14, Smith16}
and problems of MBL
or information spreading 
in the presence of power-law interactions \cite{Burin06, Yao14, Burin15, Hauke15, Li16, Gutman17, Singh17, Nandkishore17, Tikhonov18, Safavi19, DeTomasi18, Botzung18, Roy19, Schiffer19, Nag19, Kloss20, Deng20, Luitz19, Chen19,Guo20}.
 Understanding the relation of POLFED to alternative eigensolvers 
 \cite{Bollhofer07, Polizzi09, Beeumen20} is an interesting task for a further research.

 POLFED allowed us to study MBL transition in $J_1$-$J_2$ model of size $L\leqslant 24$. Such a system size is
 sufficient to demonstrate the breakdown of the scaling $t_{Th} \propto L^2 \mathrm{e}^{W/\Omega}$ of Thouless time \cite{Suntajs19}.
 Studying the system size scaling of entanglement entropy $S_E$ of eigenstates 
 we estimated the critical disorder strength of transition to MBL phase.

 {\it Acknowledgments. } 
We thank Fabien Alet and Dominique Delande for insightful discussions.
The computations have been performed within PL-Grid Infrastructure, 
its support is acknowledged. M.L. acknowledges the Spanish Ministry MINECO (National Plan
15 Grant: FISICATEAMO No. FIS2016-79508-P, SEVERO OCHOA No. SEV-2015-0522, FPI), European
Social Fund, Fundació Cellex, Fundació Mir-Puig, Generalitat de Catalunya (AGAUR Grant No. 2017 
SGR 1341, CERCA/Program), ERC AdG NOQIA, EU FEDER, MINECO-EU QUANTERA MAQS (funded by The State 
Research Agency (AEI) PCI2019-111828-2 / 10.13039/501100011033) , and the National Science Centre,
Poland-Symfonia Grant No. 2016/20/W/ST4/00314. The support of National Science Centre, Poland
under Unisono Grant No.  2017/25/Z/ST2/03029 (Quantera: QTFLAG) is also acknowledged (J.Z.).
P.S. acknowledges National Science Centre, Poland: ETIUDA grant No. 2018/28/T/ST2/00401 .

\normalem
%


\clearpage

\section{Supplementary Material}

\subsection{Convergence of POLFED}

\begin{figure}[h]
\includegraphics[width=0.475\linewidth]{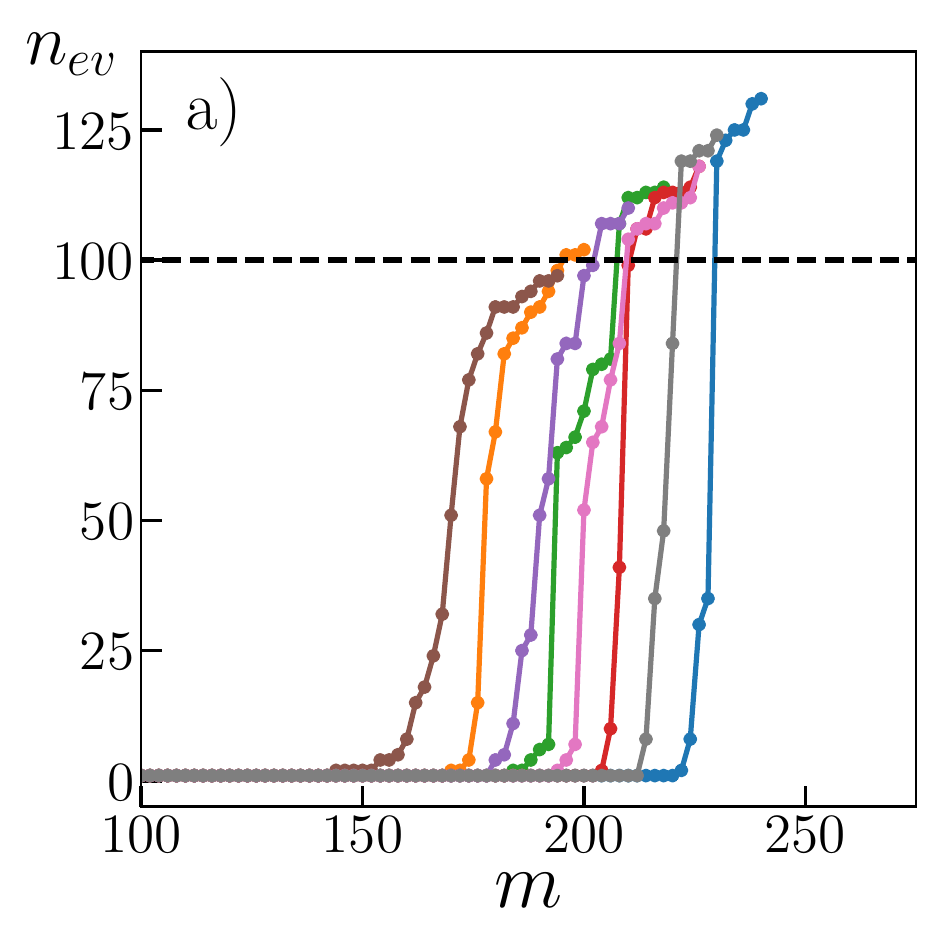}\includegraphics[width=0.48\linewidth]{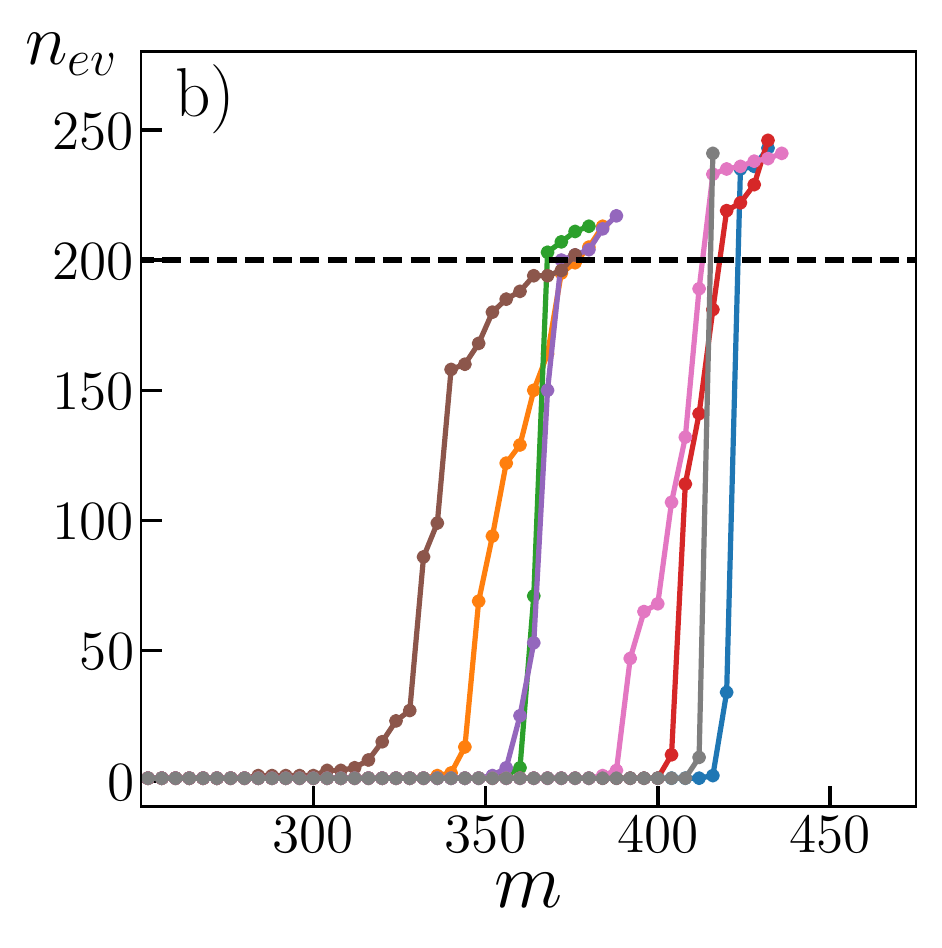} 
\includegraphics[width=0.5\linewidth]{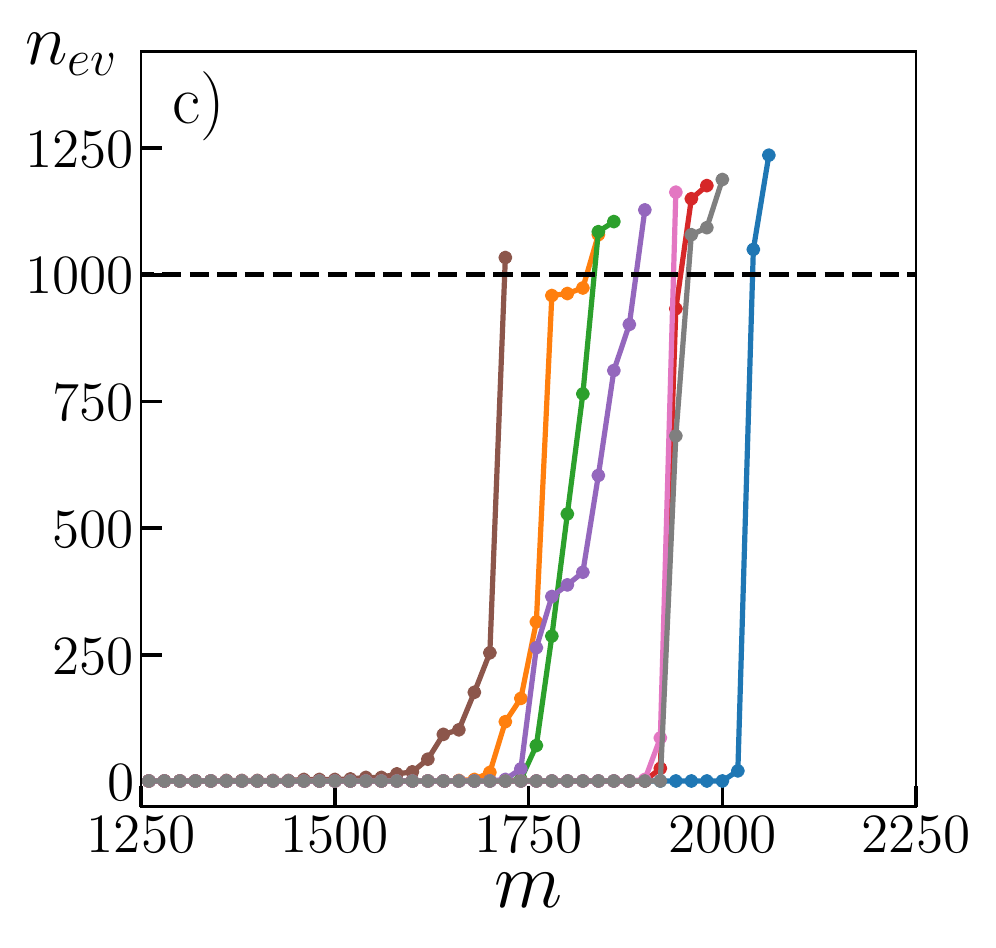}\includegraphics[width=0.47\linewidth]{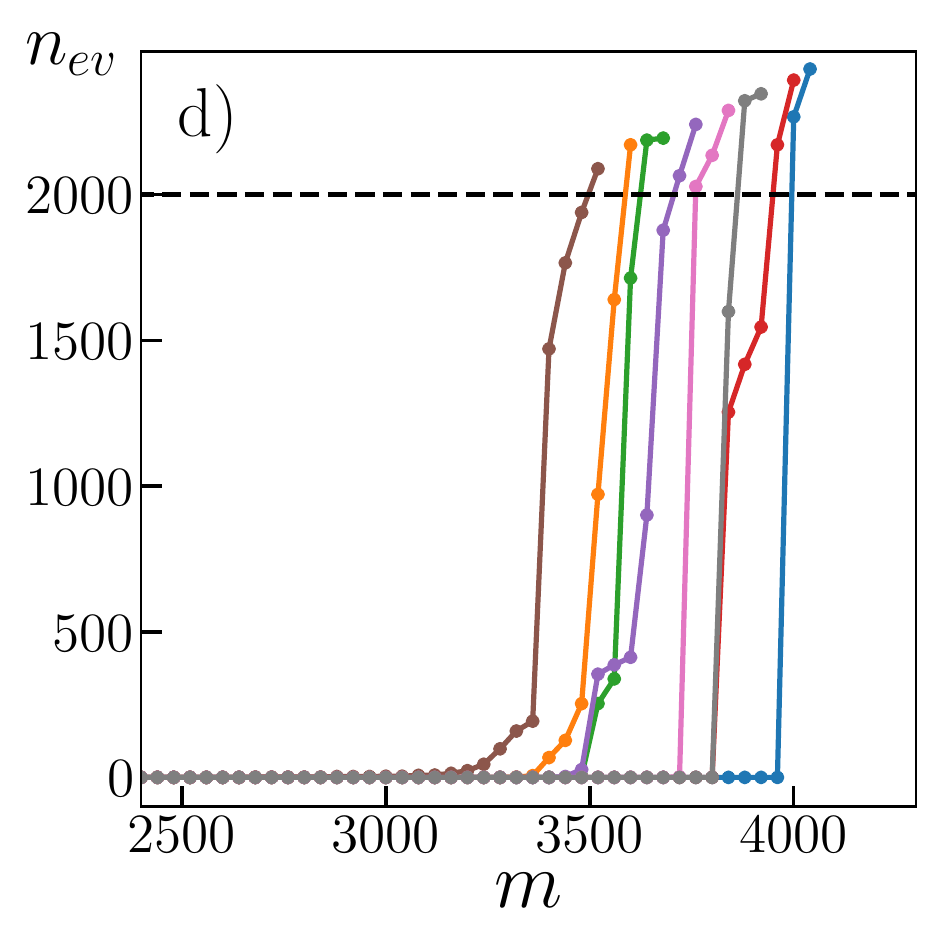} 
  \caption{ Convergence of POLFED. Number of converged eigenpairs $n_{ev}$ 
  as a function of number of Lanczos steps $m$. Dotted lines correspond to $8$
  different disorder realizations with disorder strength $W=5.4$ for   
  $J_1$-$J_2$ model, system size is $L=20$. The block size $s=1$. Panels a), b), c) and d) correspond, respectively,
  to the number of requested eigenvalues $N_{ev} = 100,200,1000, 2000$ (denoted by black dashed lines).
 }\label{convm}
\end{figure}

POLFED performs the iteration of the block Lanczos method for the 
transformed matrix $P^K_{\sigma}(\tilde H)$ until all of the residual norms
associated with  eigenvalues  $\theta_i \geqslant p=0.17$
vanish within numerical precision.
Fig.~\ref{convm} shows the number $n_{ev}$ of converged
(with the vanishing residual norm) eigenpairs 
in few runs of POLFED.
Regardless of the number $N_{ev}$ of requested eigenvalues, $n_{ev}$ 
increases rapidly, once the number $m$ of 
 Lanczos steps exceeds a certain threshold value which is typically twice larger 
than $N_{ev}$. Therefore, the eigenpairs start to converge only when the 
$\mathcal Q_m$ matrix contains the full basis of subspace of Hilbert space spanned by 
eigenvectors with eigenvalues close to the energy $\sigma$. It is beneficial to stop the algorithm
once $n_{ev} \geqslant N_{ev}$. Further increase of $m$ does not lead to an increase in $n_{ev}$ because
of the large density of states of $P^K_{\sigma}( \tilde H)$ close to $\theta\approx0.15$
due to the the secondary minima of $P^K_{\sigma}(\epsilon)$ (see Fig.~1 of the main text). 

POLFED follows the convergence pattern described above provided that the polynomial spectral transformation
$P^K_{\sigma}$ (in particular, its order $K$) is chosen in such a way that the number
$N_{ev}$ of requested eigenvalues corresponds to the number $n_P$ of eigenvalues of $P^K_{\sigma}( \tilde H )$
that are accessible to the method (i.e. fulfill the $\theta>p=0.17$ condition). To calculate $n_P$, POLFED 
uses the function $P^K_{\sigma}(\epsilon)$ as well as the density of states of the $\tilde H$. As the density of states
in the middle of the spectrum of the benchmark models considered in this work we use an analytical expression
$\tilde{\rho}(0)= (E_1-E_0)\mathcal{N}/\Gamma$,
where
\begin{equation}
 \Gamma  = \sqrt{L \left[ (1+J_2^2)/8 + \Delta^2(1+J_2^2)/16 +  W^2/12 \right]},
\end{equation}
as obtained in \cite{Suntajs19}.
The fluctuations of density of states between disorder realizations lead to the fluctuations of the threshold value 
of $m$ beyond which the convergence occurs -- see Fig.~\ref{convm}. Those fluctuations are enhanced when disorder strength 
increases. However, our tests indicate that the convergence occurs 
for each of the considered disorder values ($W\leqslant15$) and disorder realizations
for $m < 2.8 N_{ev}$. 

Testing a variety of the polynomial spectral transformation $P^K_{\sigma}( \tilde H )$ as well as different stopping criteria,
we checked that POLFED allows to minimize the time of calculation until the convergence is reached and, at the same time, 
allows to keep a relatively large the total number of eigenpairs obtained in a single run.

When the block size $s$ of the Lanczos method is increased, the total number of vectors generated in the iteration, $ms$,
required for the convergence of algorithm, is also increased. However, for the typical production runs
done in this work, i.e. with 
the block size $s \leqslant 24$ and $N_{ev} \geqslant 1000$, the total number of Lanczos vectors still fulfills
the condition $ms < 2.8 N_{ev}$. Thus, during its start, POLFED allocates $2.8N_{ev}$ columns of the matrix $\mathcal Q_m$. 
The associated memory consumption is proportional to $N_{ev} \mathcal N$, which has the dominant contribution to total memory occupation 
of POLFED.

\subsection{Technical details of POLFED}

Calculation of the product of $P^K_{\sigma}( \tilde H)$ with subsequent columns of $Q_j$
is the most time consuming step of POLFED. 
The recurrence relation $T_{n+2}(x)=2 x T_{n+1}(x) - T_{n}(x)$
fulfilled by Chebyshev polynomials reduces this product to multiplication
of vectors by the sparse matrix $ \tilde H$ and basic linear algebra operations.
The Clenshaw algorithm  \cite{Clenshaw55}, allows us to 
reduce the number operations needed to calculate the product.

The efficiency of computation of $P^K_{\sigma}( \tilde H)x$ where $x \in \mathbb R^{\mathcal N}$ 
is crucially dependent on efficiency of the single sparse matrix vector multiplication $\tilde H x$.
In the current version of POLFED we store the $\tilde H$ matrix in CSR format. 
We do not store the off-diagonal Hamiltonian entries of $H$ as they are all equal to $H_{ij}=1/2$.
On one hand this reduces the memory
consumption associated with storing of the Hamiltonian matrix. On the other hand, POLFED does not access
the values of $H_{ij}$ during the matrix-vector multiplication which increases the efficiency of the code.
Throughout this work, we 
consider block sizes $s \leqslant 24$ calculating the products of  $P^K_{\sigma}( \tilde H)$ 
with columns of matrix $Q_j$ independently. Each of the products is calculated on a single core. Effectively, POLFED 
performs the computation in parallel on $s$ cores. 
The re-orthogonalization of columns of matrix $U_j$ obtained in Lanczos step against the columns of 
matrices $ \{Q_i\}_{i=1}^j$ is parallelized similarly: each of the $s$ cores orthogonalizes a single 
column of $U_j$ against the columns of $\{Q_i\}_{i=1}^j$.

The matrix-vector multiplications $\tilde H x$ could be performed on multiple cores with use of external 
sparse basic linear algebra libraries, resulting in higher degree of parallelism in POLFED. 
Moreover, the promising way of enhancing the performance of POLFED is to optimize the sparse matrix-vector
product, a subject that recently received attention both on CPUs as well as on GPUs
\cite{Bell08,Acer16,  Chen18A, Baca19}.

\subsection{Benchmark for disorder spin chains}

In this section we compare performance of POLFED with state-of-the-art SIMED code for $XXZ$ and $J_1$-$J_2$
models. Benchmark results are shown in Tab.~\ref{tabXXZ} and in Tab.~\ref{tabJ1J2}.

The linear scaling of 
density of states $\rho(0)$ with $\mathcal N$ implies that the order of the polynomial spectral transformation $K \propto \mathcal N$. Therefore, up to a factor polynomial in $L$, the computation time of POLFED scales as $\mathcal N^2$. The total CPU time $t_{CPU}$ for POLFED increases by a factor of $\approx16$
 both for $XXZ$ model (Tab.~\ref{tabXXZ})
and for $J_1$-$J_2$ model (Tab.~\ref{tabJ1J2}) when the system size $L$ increases by $2$.
Typically, the increase of $t_{CPU}$ is slightly larger when the number $N_{cores}$ of cores
(equal to the block size $s$ for POLFED) increases. The memory consumption of POLFED indeed scales as $N_{ev} \mathcal N$ up to constant a additive factor
due to the storing of the Hamiltonian matrix as Tab.~\ref{tabXXZ} and Tab.~\ref{tabJ1J2} show. 
The low memory consumption of POLFED allows for calculations on a single node ($N_{cores}\leqslant 24$ on
the supercomputer Prometheus, ACK Cyfronet AGH, Krakow)
 for both models as long as $L \leqslant 24$.

\begin{table}
\begin{tabular}{c c S c c c c c c } \toprule
 &  & $L\,\,\,\,\,\,\,$   & {$t_{CPU}[h]$} & {$N_{cores}$} & {$ t_{W}[h]$}  
& {$RAM [GB]$} & {$  N_{ev} $} \\ \midrule
\parbox[t]{3mm}{\multirow{3}{*}{\rotatebox[origin=c]{90}{POLFED}}} &  &    20  & 3.1 & 1 & 3.1 & 3.9 & 1000 \\
  &  & 22  & 62.2 & 4 & 15.5 & 21.2 & 1400  \\
  & & 24  &  1503 & 24 & 62.6   & 114 & 2000\\
 &  & 26  &    19870 & 24 & 828 &  488 & 2000\\  \midrule
 \parbox[t]{3mm}{\multirow{3}{*}{\rotatebox[origin=c]{90}{SIMED $\,$}}} & &  20  & 0.5 & 20 & 0.026 &  22 & 100\\
 &  & 22  & 20.2 & 120 & 0.17 & 244 & 100  \\
 &  & 24  &  840 & 2880 &  0.23 & 12288 & 50\\
 &  & 26  &  36000 & 48000 & 0.75 &  204800 & 50\\     \bottomrule
\end{tabular}
 \caption{ \label{tabXXZ} {
 POLFED vs SIMED for $XXZ$ spin chain:
 $t_{CPU}$ is total CPU time,  $N_{cores}$ is the number of cores used in calculation, $t_W$ 
 is total execution time, $RAM$ is total memory
occupation, $N_{ev}$ is the number of obtained eigenpairs in the middle of the spectrum ($\sigma=0$).
Tested on 
Intel Xeon E5-2680v3 (2.5GHz); SIMED data for $L=20,22$ obtained on Intel Ivybridge E5-2680 (2.8GHz),
extracted from \cite{Pietracaprina18}.
}
 } 
\end{table}

\begin{table}
\begin{tabular}{c c S c c c c c c } \toprule
 &  & $L\,\,\,\,\,\,\,$   & {$t_{CPU}[h]$} & {$N_{cores}$} & {$ t_{W}[h]$}  
& {$RAM [GB]$} & {$  N_{ev} $} \\ \midrule
\parbox[t]{3mm}{\multirow{3}{*}{\rotatebox[origin=c]{90}{POLFED $\quad$}}} &  &    20  & 3.1 & 1 & 3.1 & 0.8 & 100 \\
      &  & 20  &  3.6 & 1 & 3.6 & 3.9 & 1000  \\
      &  & 22  &  60.2 & 1 & 60.2 & 3.4 & 100  \\
&  & 22  & 63.2 & 2 & 31.6 & 4.5 & 200  \\
&  & 22  & 105 & 8 & 13.1 & 21.3 & 1400  \\
  & & 24  &   3400 & 24 & 142   & 115 & 2000\\ \midrule
 \parbox[t]{3mm}{\multirow{3}{*}{\rotatebox[origin=c]{90}{SIMED}}} & &  20  & 2.4 & 36 & 0.067 & 70 & 100\\
 &  & 22  & 100 & 468 & 0.22 & 1840 & 100  \\
 &  & 22  &  120 & 468 &  0.26 & 1840 & 200\\   \bottomrule
\end{tabular}
 \caption{ \label{tabJ1J2} {
 POLFED vs SIMED for $J_1$-$J_2$ spin chain:
 $t_{CPU}$ is total CPU time,  $N_{cores}$ is the number of cores used in calculation, $t_W$ 
 is total execution time, $RAM$ is total memory
occupation, $N_{ev}$ is the number of obtained eigenpairs in the middle of the spectrum ($\sigma=0$).
POLFED tested on
Intel Xeon E5-2680v3 (2.5GHz); SIMED tested on Intel Xeon Gold 6140 CPU (2.3GHz), data provided by
courtesy of F. Alet.
}
 } 
\end{table}
 The memory consumption of SIMED is dominated by the factors obtained in $LU$ decomposition of the Hamiltonian,
 it scales as $c(L) \mathcal N$ (up to terms polynomial in the system size $L$). 
 The factor $c(L)$ describes fill-in of the matrix. Tests performed in 
 \cite{Pietracaprina18} indicate a phenomenological scaling $c(L) \propto 3^{L/2}$ for $XXZ$ spin chain. 
 This results in total memory needed to 
 store the $LU$ factors to be $\approx 2000GB$ and $14000 GB$ respectively for $L=24$ and $L=26$. The actual memory
 usage, due to peaks of allocated/de-allocated memory in the SIMED is significantly higher
 as shown in Tab.~\ref{tabXXZ}. The rapid increase of memory consumption with $L$ 
 forces one to use a very large number of nodes in calculations with SIMED, eventually making the calculations infeasible,
 even on large supercomputers. Theoretically, the calculation time of SIMED, dominated by the $LU$ factorization, should 
 be proportional to the number of elements in the factors yielding the scaling of total CPU time 
 $t_{CPU} \propto c(L) \mathcal N$. However, as Tab.~\ref{tabXXZ} shows, $t_{CPU}$ scales more rapidly with system size,
  increasing approximately $40$ times when $L$ increases by $2$. Altogether, the system size scaling of
  $t_{CPU}$ is better for POLFED.

  Another aspect of the fill-in phenomenon of SIMED is that it is quite unpredictable. For instance, 
  the coefficient $c(L)$ may change after reordering of the basis. It is, however, clear that the fill-in 
  becomes much more severe as the number $N_{nz}$ of non-zero off-diagonal elements increases. Tab.~\ref{tabJ1J2}
  shows that the total memory consumption of SIMED for $J_1$-$J_2$ model
  is increased, in comparison to resources needed for $XXZ$ model,
  by a factor of $\approx 3.5$ and $\approx7.5$ respectively for $L=20$ and $L=22$.
  The total CPU times $t_{CPU}$ for POLFED and SIMED for $J_1$-$J_2$ model are very similar for $L=20, 22$. However,
  the rapidly increasing memory usage of SIMED makes the calculations for $L=24$ infeasible on present day supercomputers.
  At the same time, POLFED allows to obtain results for $J_1$-$J_2$ model of size $L=24$ with resources similar to the $XXZ$ model --
  such a calculation fits in a single node of a supercomputer.
  
  Another advantage of POLFED is that it allows for a substantial increase of the number of requested eigenvalues,
  $N_{ev}$, without a significant increase in the total calculation time. This can be readily understood. 
  When replacing $N_{ev} \rightarrow \alpha N_{ev}$ where $\alpha>1$, the condition that $N_{ev}$ eigenvalues 
  of $P^K_{\sigma}( \tilde H)$ are larger than $p=0.17$ results in the order of the spectral transformation
  $K\rightarrow K/\alpha$. Even though the total number of Lanczos iterations needed for the convergence of the 
  algorithm increases by a factor of $\alpha$, the cost of calculation of a single polynomial spectral transformation
  decreases $\alpha$ times. The re-orthogonalization performed by POLFED is the only source of increase of total 
  CPU time when $N_{ev} \rightarrow \alpha N_{ev}$. This can be seen in Tab.~\ref{tabJ1J2} as tests for $L=20,22$ 
  were performed for few values of $N_{ev}$.
  The change of total CPU time with number of requested eigenvalues is more significant for SIMED. For instance, Fig.~6. of 
  \cite{Pietracaprina18} shows that increase of $N_{ev}$ from $100$ to $1000$ results in approximately $6$ times
  larger $N_{ev}$ for $XXZ$ model of size $L=20$.

 Typically, MBL calculations require averaging over disorder realizations. The POLFED allows to find eigenpairs
 of the disordered spin chains on relatively small number of cores so that the averaging over disorder 
 realizations can be done by performing calculations for many disorder realizations independently
 at the same time. In contrast, SIMED requires much larger amount of resources. Ultimately, due to smaller total execution times,
 for $XXZ$ model 
 of sizes $L\leqslant 24$ SIMED allows to get results for a comparable, but slightly larger number of 
 disorder realizations using a fixed amount of CPU time (assuming than one is able to perform calculations
 for $L=24$ simultaneously on $120$ nodes of a cluster). For $J_1$-$J_2$ model at system size $L=20,22$ POLFED has an advantage.
 Moreover, while SIMED calculations for $J_1$-$J_2$ model at $L=24$ are infeasible, they can be readily done 
 by POLFED. The situation is similar for $XXZ$ model at $L=26$: single SIMED run requires $2000$ nodes of a supercomputer,
 and the calculation is performed in single precision \cite{Pietracaprina18} whereas POLFED calculation, in double precision,
 requires only few nodes of a supercomputer.

\subsection{Extraction of Thouless time}

To extract Thouless time $t_{Th}$ from spectral form factor $K(\tau)$ for system size $L \leqslant 18$
we use data from full exact diagonalization and follow the procedure
outlined in \cite{Suntajs19}.

To this end we calculate the spectral form factor (SFF) according to its definition
      \begin{eqnarray}
 K(\tau)  = \frac{1}{Z}\left \langle \left| \sum_{j=1}^{\mathcal N} g( E_j) \mathrm{e}^{-i E_j \tau} \right|^2 \right \rangle.
 \label{eq: KtS}
\end{eqnarray}
Subsequently, we perform the unfolding, during which the  
level staircase function $\sigma( \epsilon )= \sum_i \Theta(\epsilon- \varepsilon_i)$ 
(obtained from the set of eigenvalues of the system $\{ \varepsilon_i\}$ ordered in ascending manner)
is separated into smooth and fluctuating parts $\sigma(E) = 
\overline \sigma(E) + \delta \sigma (E)$ and the eigenvalues are mapped via 
 $ \varepsilon_j \rightarrow E_j = \overline \sigma( \varepsilon_j)$.
As the smooth part $\overline \sigma(E)$ we take a polynomial of degree $n_p=10$ 
fitted to the level staircase function $\sigma (E)$.
To calculate SFF we use
$g(\epsilon) \propto \exp( -(\epsilon -\bar \epsilon)^2/{2 \eta \sigma_{\epsilon}}^2 )$,
where $\bar \epsilon$ denotes the average of the unfolded eigenvalues for given disorder
realization $\epsilon_i$, $\sigma_{\epsilon}$ is the standard deviation of $\{ \epsilon_i\}$ 
and $\eta = 0.3$. This choice of parameters follows precisely \cite{Sierant20}.
Then, we calculate
 \begin{equation}
  \Delta K(t/t_H) = \left| \log\left( \frac{K(t/t_H)}{K_{GOE}(\tau=t/t_H)}\right)\right|,
  \label{DeltaK}
 \end{equation}
 where the spectral form factor for GOE is given by
\begin{equation}
\label{fitKt}
K_{GOE}(\tau) =
\begin{cases}
2 \tau - \tau \ln(1+2\tau)  \quad \mathrm{for} \quad \tau \leqslant 1,\\ 
2 - \tau \ln( \frac{2\tau+1}{2\tau-1} ) \quad \quad \, \mathrm{for} \quad \tau > 1.
\end{cases}
\end{equation}
 The Thouless time $t_{Th}$ is the smallest positive time for which $\Delta K(t/t_H)< a$.
 We choose the value of cut-off $a = 0.1$. 

 \begin{figure}[h]
  \includegraphics[width=0.99\linewidth]{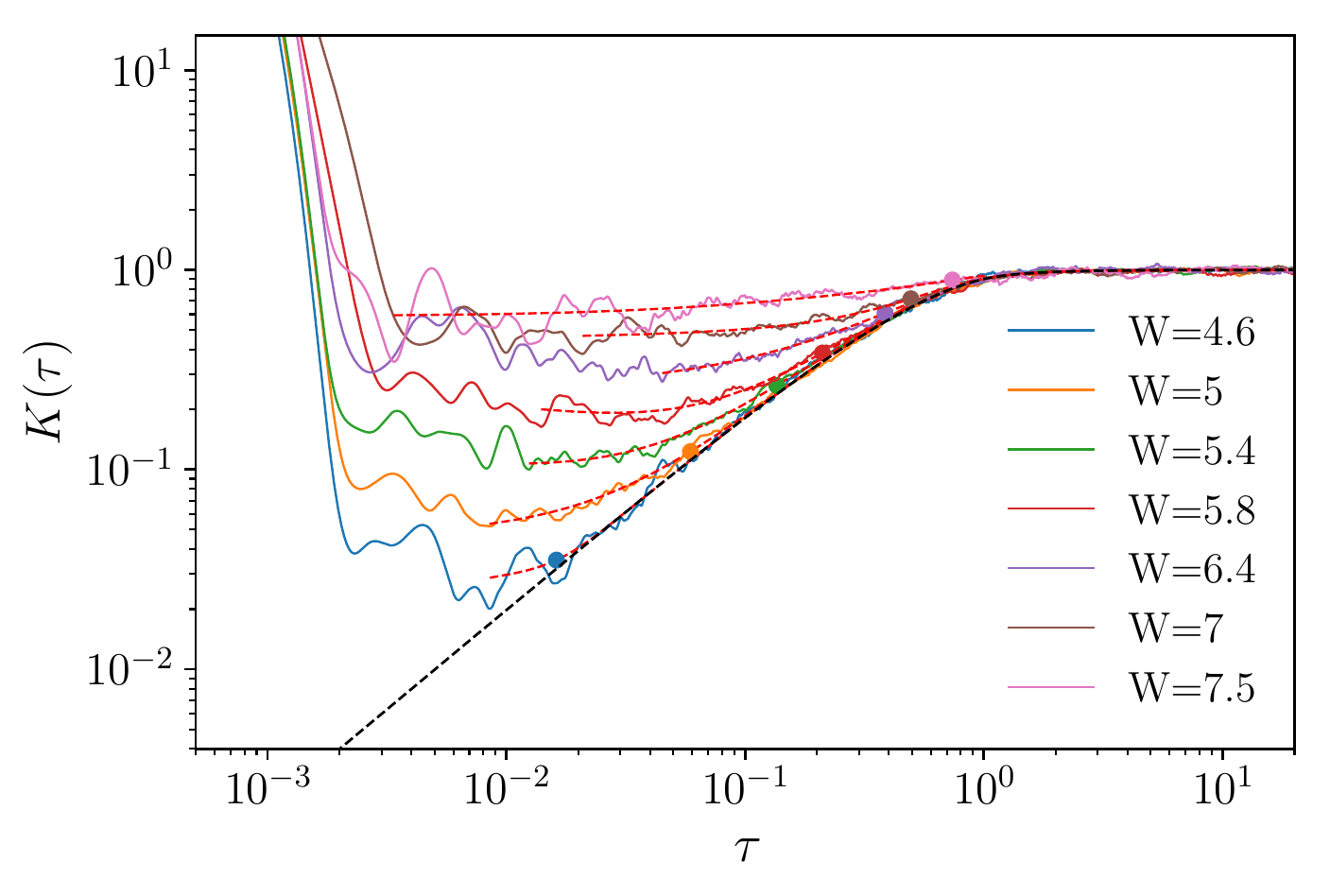} 
  \caption{ Spectral form factor $K(\tau)$ of $J_1$-$J_2$ model, system size $L=24$. 
  Black dashed lines shows spectral form factor of GOE, $K_{GOE}(\tau)$. The dashed  
  lines show $K_F(\tau)$ \eqref{fitKt} fitted  to $K(\tau)$; dots denote the obtained rescaled Thouless time:
  $\tau_{Th} = t_{Th} / t_{H}$, where $t_{H}$ is the Heisenberg time.
 }\label{Ktfig}
\end{figure}

For system sizes $L=20,22,24$ we obtain $N_{ev}=2500$ consecutive eigenvalues from the middle of spectrum.
Firstly, we perform the unfolding procedure using fitting the level staircase function with a polynomial of degree 
$n_p'=3$. Then, we calculate the spectral form factor according to the definition \eqref{eq: KtS} considering only
the calculated eigenvalues in the sum. Since the number of disorder realizations we have for the largest system 
considered ($L=24$) is only $50$, we fit the spectral form factor with the following formula
\begin{eqnarray}
\label{KtFIT}
K_F(\tau) = K_{GOE}(\tau) + c_1  \exp \left(-c_2 \tau^{c_3} \right),
\end{eqnarray}
where $c_1, c_2, c_3$ are fit parameters. The results are shown in Fig.~\ref{Ktfig}. The formula \eqref{KtFIT} provides very good fits of 
to the spectral form factor for smaller system sizes ($L\leqslant 20$) for $\tau \lesssim 0.2 \tau_{Th}$.
Thus, to extract $\tau_{Th}$ for $L=22, 24$ we use $K_F(\tau)$ in \eqref{DeltaK}.

Fig.~\ref{Ktfig} illustrates also an another aspect of calculation of the Thouless time when not all of the eigenvalues of the system
are available. The number of $N_{ev}$ eigenvalues determines the value of $\tau_{N_{ev}}$ below which the spectral form factor 
rapidly increases. In our case, as can be seen in Fig.~\ref{Ktfig}, $\tau_{N_{ev}} \approx 2 \cdot10^{-3}$. Once the extracted value of 
$\tau_{Th}$ is significantly bigger than $\tau_{N_{ev}}$, the value of $\tau_{Th}$ is not affected by the fact that $N_{ev} \ll \mathcal N$.

\subsection{ Extraction of $W^*$}

\begin{figure}
\includegraphics[width=1\linewidth]{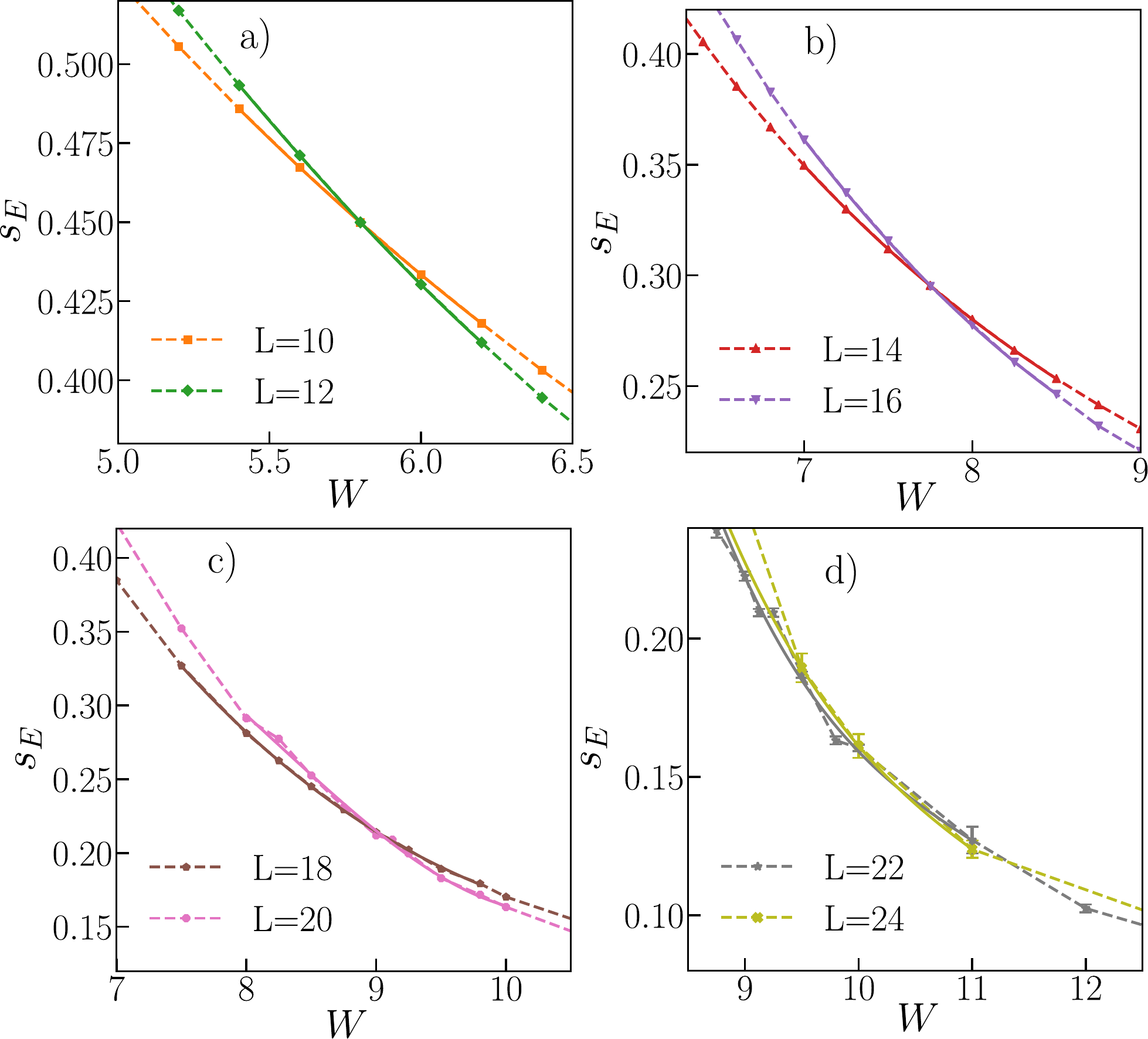}
  \caption{ Determination of crossings $s_E(L-1)=s_E(L+1)$ for $J_1$-$J_2$ model. The rescaled
  entropy $s_E$, denoted by dashed line, is plotted as a function of disorder strength $W$ for system sizes $L=10,12$,
  $L=14,16$, $L=18,20$, $L=22, 24$ respectively on panels a), b), c), d). $s_E(W)$ curves are fitted
  with polynomials of third degree (shown by solid lines) in vicinity of the crossing point.
  The crossing points of the polynomials determine $W^*_E(L)$ for $L=11,15,19,23$.
 }\label{j1j2DET}
\end{figure}
To extract the values of $W^*_E{(L)}$ we plot (for odd $L$), $s_E(L-1)$ and $s_E(L+1)$ as functions of disorder strength
$W$, and perform a fit with third order polynomial in the vicinity of the crossing point, examples are shown in
Fig.~\ref{j1j2DET}. The crossing point of the two polynomials is then the value of $W^*_E{(L)}$. A similar, 
analysis is performed for even $L$ and data for $s_E(L-2),\, s_E(L+2)$.

The values of $W^*_{\overline r}{(L)}$ are found in analogous manner from $\overline r(L-1)$ and $\overline r(L+1)$
($\overline r(L-2)$ and $\overline r(L+2)$) as functions of disorder strength $W$ for odd (even) $L$. 

\begin{figure}
  \includegraphics[width=0.99\linewidth]{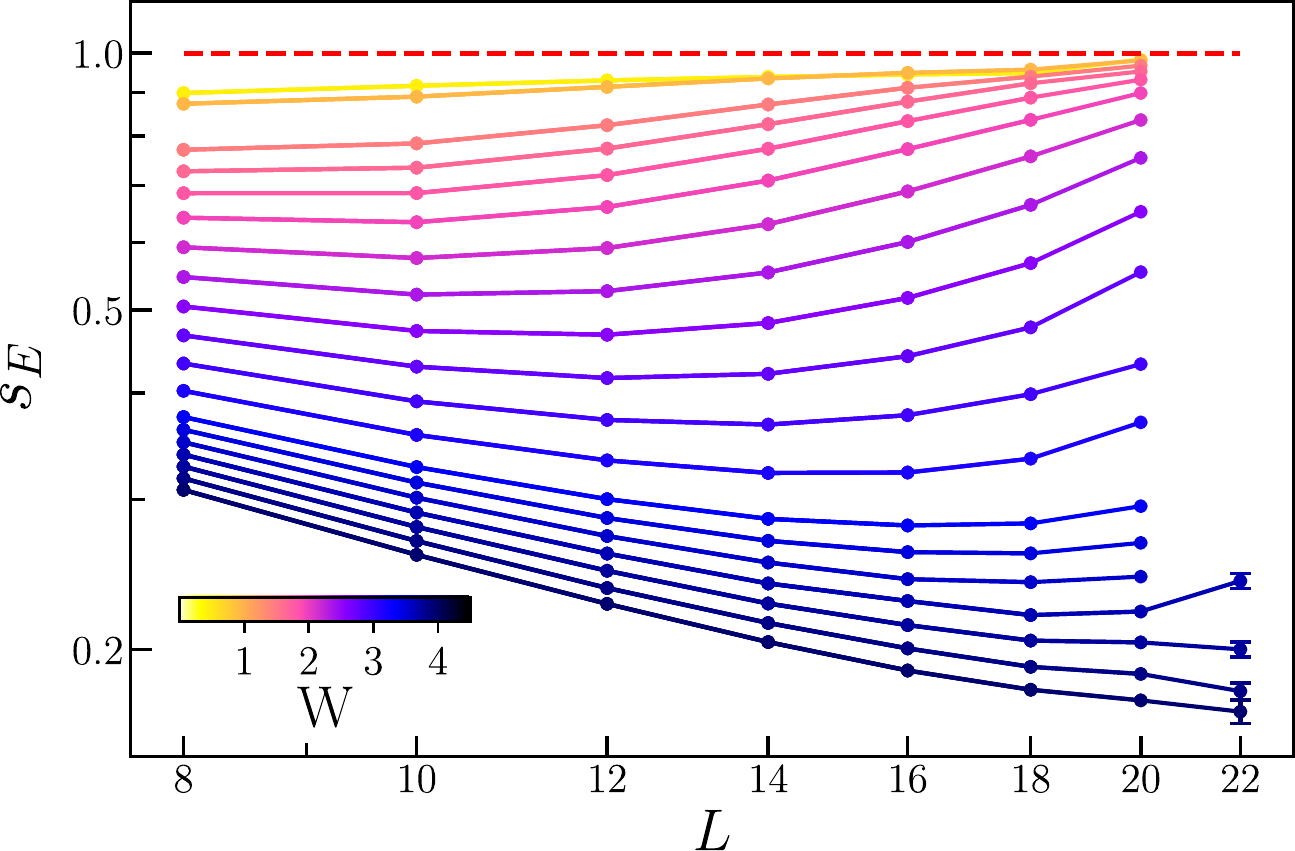} 
  \caption{The rescaled entanglement entropy $s_E$
  of eigenstates of $XXZ$ model as a function of system size $L$ 
for disorder strengths  $W=0.5,..., 4$ (denoted on the color bar). Dashed lines correspond to ergodic  behavior
$s_E=1$.
 }\label{entXXZ}
\end{figure}

\begin{figure}
  \includegraphics[width=0.99\linewidth]{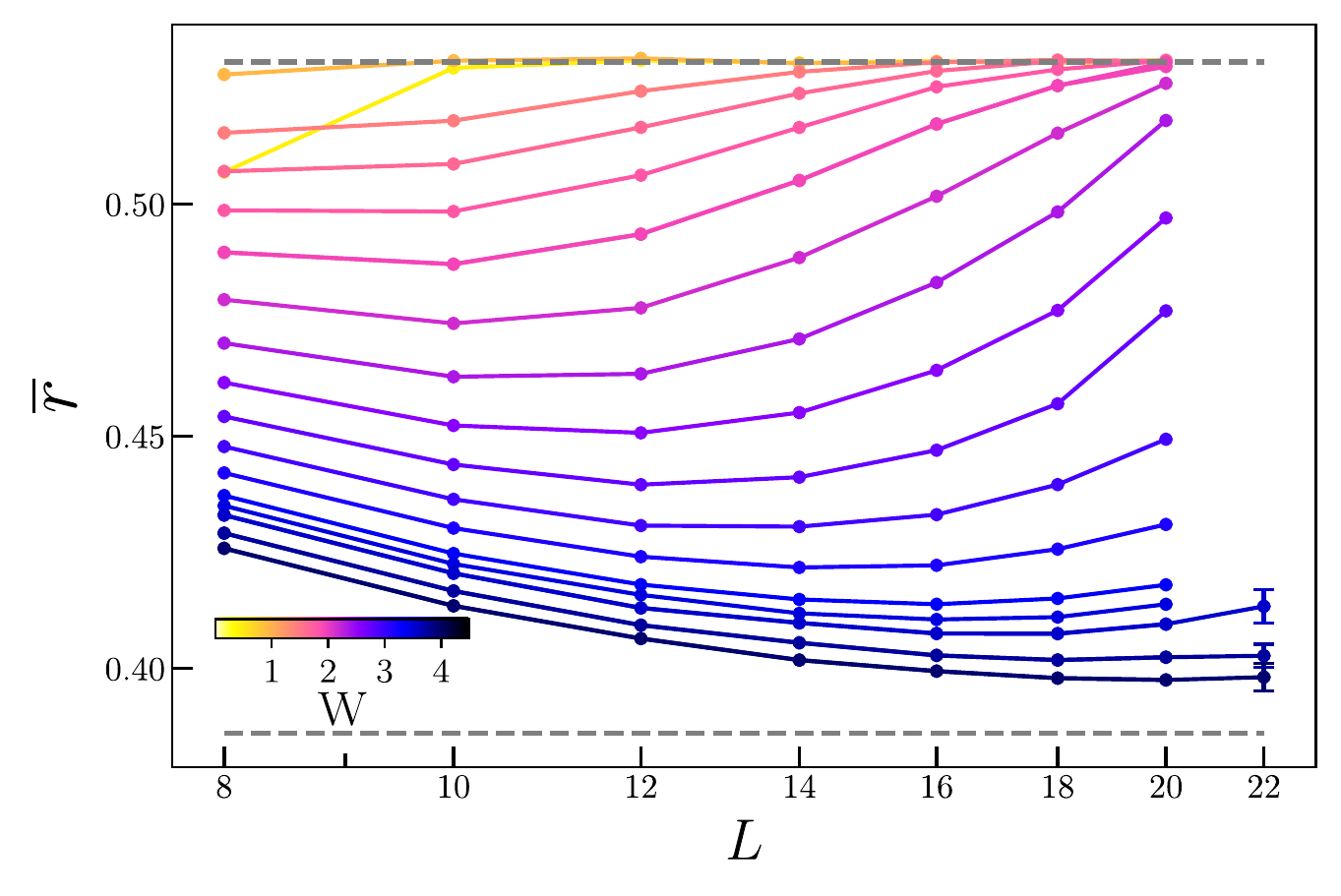} 
  \caption{ The average gap ratio $\overline r$
  for $XXZ$ model as a function of system size $L$ 
for disorder strengths  $W=0.5,..., 4$ (denoted on the color bar). Dashed lines correspond to ergodic  behavior
$s_E=1$.
 }\label{rbarXXZ}
\end{figure}

We perform a similar analysis for $XXZ$ model obtaining $W^*_{E, \overline r}{(L)}$ for that system. The scaled entanglement entropy and
average gap ratio for $XXZ$ model are shown in Fig.~\ref{entXXZ} and Fig.~\ref{rbarXXZ}.
The scaled entanglement entropy $s_E$ is obtained when 
we average $S_E$ over the position of the cut $x$, over 
$N_{ev} \leqslant \min\{\mathcal N/100, 2000\}$ eigenstates in the middle of the spectrum
($\sigma=0$) of $XXZ$ model for system sizes $12\leqslant L \leqslant 22$ (whereas for $L=8,10$ we take 
$N_{ev}=5$) -- similarly as for $J_1$-$J_2$ model. Then, an average over  
more than $5000$, $200$ disorder realizations respectively for $L\leqslant20$, $L=22$ is performed. Eigenvalues corresponding 
to eigenstates used in calculation of $s_E$ are employed in computation of the average gap ratio $\overline r$.


\end{document}